\documentclass[journal=nalefd,manuscript=articlen]{achemso}

\usepackage{chemformula} 
\usepackage[T1]{fontenc} 
\usepackage{amsmath}
\usepackage{graphicx}
\usepackage{txfonts}
\usepackage{subcaption}
\usepackage{multirow}
\author{Mohsen Salimi}
\author{Robin V. Nielsen}
\author{Henrik B. Pedersen}
\author{Aur\'{e}lien Dantan}
\email{dantan@phys.au.dk}
\affiliation{Department of Physics and Astronomy, University of Aarhus, DK-80000 Aarhus C, Denmark}

\title[]{Squeeze film absolute pressure sensors with sub-millipascal sensitivity}

\abbreviations{}

\begin{document}

\begin{tocentry}

\end{tocentry}

\begin{abstract}
We report on the realization of ultrasensitive absolute pressure sensors based on silicon nitride membrane sandwiches. These sandwiches consist in a pair of highly-pretensioned,  ultrathin (50 nm), large area (0.25 mm$^2$) films, suspended parallel to each other and forming an ultrashort (500 nm), open cavity. The compression of a gas in this cavity leads to a strong squeeze film force, resulting in an increase in the membrane mechanical resonance frequencies which is directly proportional to the absolute gas pressure. These sandwiches show a record high responsitivity of $>300$~Hz/Pa in terms of squeeze film-induced frequency shift, which, combined with high quality factor mechanical resonances ($Q>10^6$), allows for bringing the sensitivity of absolute squeeze film pressure sensors down to the sub-millipascal level.
\end{abstract}


Absolute pressure measurements in high or ultrahigh vacuum are notoriously challenging, as they typically require a very sensitive measurement of the mechanical force exerted by a gas on a mechanically compliant or vibrating structure. In conventional sensors the pressure difference with a reference cavity can be determined by measuring the deflection of a flexible diaphragm or by monitoring the change in its mechanical resonance frequencies~\cite{Bao2000}. While such ``pressure-difference" sensors can be highly sensitive~\cite{Zhang2001,AlSumaidae2021,Chen2022}, especially when relying on miniaturized cavities and high-mechanical quality nanodiaphragms, the fabrication of sealed, long term-stable, outgassing-free microcavities is generally a challenge. 

{\it Squeeze film} sensors, which exploit the compression of the gas in a small, open cavity to alter the mechanical properties of a beam or a diaphragm, bypass these sealed cavity sensor issues and offer a promising and potentially technically simpler alternative. Diaphragms with various geometries, materials, and readout (resistive, capacitive, optical) methods have been applied to squeeze film pressure measurements in low pressure environments~\cite{Prak1991,Andrews1993,Steeneken2005,Mol2009,Suijlen2009,Southworth2009,Suijlen2012,Kainz2014,Kumar2015,Dolleman2016,Naesby2017,Naserbakht2019,Siskins2020}. A particularly interesting avenue is the combination of low mass diaphragms diplaying high frequency and high Q resonances---as e.g. graphene microdrums~\cite{Dolleman2016}, for which impressive pressure responsivities of 90 Hz/Pa have been reported for mechanical frequencies in the 10 MHz range---with noninvasive optical readout methods, which allow for preserving their mechanical properties.

Other diaphrams which have been successfully applied to squeeze film pressure sensing are highly-pretensioned silicon nitride (SiN) membranes, which possess large area (mm$^2$) and low thickness (tens of nanometers) together with high Q (>10$^6$) resonances in the few hundreds of kilohertz to a few megahertz range. Assembling these robust, commercially available membranes into sandwiches makes it possible to realize small gap, open cavities in which the squeeze film force is essentially elastic~\cite{Naesby2017,Naserbakht2019,Dantan2020}. This results in an added pressure-dependent stiffness for the mechanical resonator, which, in the case of an isothermal compression, changes its mechanical resonance frequency $f$ by an amount
\begin{equation}
\Delta f=\frac{P}{8\pi^2 f\rho t d},
\label{eq:shift}
\end{equation}
where $P$ is the ambient pressure, $\rho$ the membrane density, $t$ its thickness, and $d$ the gap between the two parallel membranes. This expression is valid in the molecular flow regime, for membranes whose lateral dimension is much larger than the gap and whose oscillation period is much shorter than the pressure equalization time, i.e. the time it takes gas molecules to escape the gap region. It is also valid only for small frequency shifts, $\Delta f\ll f$ (see e.g.~\cite{Dantan2020} for further details). The pressure responsivity in terms of resonance frequency shift, $(\Delta f/P)$, is gas-independent and obviously increases for thin membranes with low frequency resonances as well as for small gaps. 

In addition to this squeeze-film added stiffness, the membranes experience an added {\it kinetic} damping due to collisions of the gas molecules~\cite{Christian1966}
\begin{equation}
\gamma_\textrm{kin}=4\sqrt{\frac{2M}{\pi RT}}\frac{P}{\rho t},
\label{eq:gammakin}
\end{equation}
where $M$ is the gas molar mass, $R$ the ideal gas constant and $T$ the temperature. While the relation between the kinetic damping and the pressure can also be used to sensitively determine the ambient pressure~\cite{Reinhardt2023}, it requires knowledge of the mass of the molecules (or the gas composition in case of a mixture). Depending on the resonator material, it may also require knowledge of the accomodation coefficient, i.e. the probability that a molecule sticks to the surface after a collision, which will typically be gas- and pressure-dependent in a non-trivial fashion~\cite{Zhao2023}. 

In addition to the kinetic damping the membranes may also experience a squeeze film damping due to the finite response time of the gas to a pressure change. Denoting by $\tau$ the pressure equalization time, i.e. the time it takes for the gas molecules to leave/enter the gap region, the squeeze film damping is, in the elastic response regime characterized by a large Weissenberg number, $W_i=2\pi f\tau\gg 1$, given by ~\cite{Suijlen2009}
\begin{equation}
\gamma_\textrm{sq}=\frac{P}{4\pi^2f^2\rho t d}\frac{1}{\tau}.
\label{eq:gammasq}
\end{equation}
The pressure equalization time will depend on the mass of the gas, the geometry of the microstructure and the nature, elastic or inelastic, of the collisions between the membranes and the gas molecules, and can be difficult to access experimentally. In contrast, the squeeze film-induced frequency shift of Eq.~(\ref{eq:shift}) provides a direct and robust way to absolute pressure determination.

Previous SiN sandwiches with 100 nm-thick membranes and gaps in the 2-3 $\mu$m range showed responsivities in terms of squeeze film-induced frequency shift at the 30 Hz/Pa level for 800 kHz mechanical resonances with $Q=2\pi f/\gamma\sim 10^5$, and allowed for sensitivities at the 0.1 Pa level~\cite{Naserbakht2019}. In this Letter, we report on the realization and characterization of SiN membrane sandwich squeeze film pressure sensors with a 10-fold enhanced responsivity and 100-fold enhanced sensitivity, considerably improving the performances of squeeze film pressure sensors and bringing their sensitivity at the level of the best commercial capacitive drum sensors.\\

\begin{figure}
\begin{subfigure}{0.29\columnwidth}
\centering
\includegraphics[width=\columnwidth]{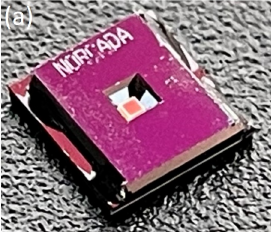}
\end{subfigure}
\begin{subfigure}{0.29\columnwidth}
\centering
\includegraphics[width=\columnwidth]{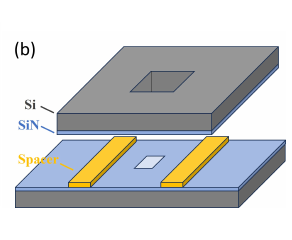}
\end{subfigure}
\begin{subfigure}{0.4\columnwidth}
\centering
\includegraphics[width=\columnwidth]{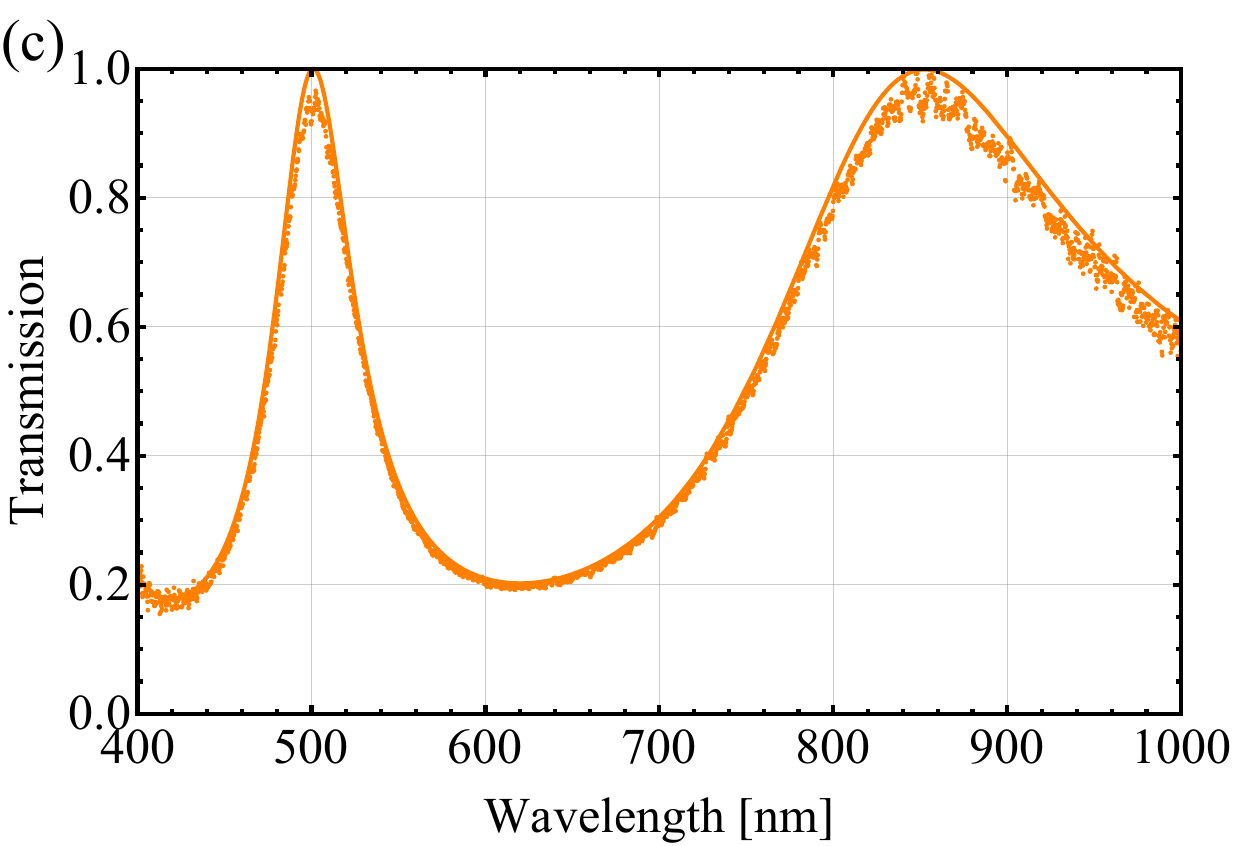}
\end{subfigure}
\caption{(a) Photograph of the membrane sandwich. (b) Sideview schematic of the two membranes prior to assembly (not to scale). (c) Normalized optical transmission of the sandwich under broadband illumination after assembly. The solid line shows the result of a fit with the transmission of a planar Fabry-Perot consisting of two lossless dielectric slabs~\cite{Nair2017}---whose thicknesses and refractive index are independently determined using ellipsometry---separated by a gap of 500 nm.}
\label{fig:trans}
\end{figure}

A photograph of such a sandwich is shown in Fig.~\ref{fig:trans}(a). The sandwich consists of a pair of highly-pretensioned ($\sim 0.3$ GPa), 500 $\mu$m-square and 55/60 nm-thick SiN membranes fabricated by Norcada~\cite{Norcada}. The 60 nm-thick membrane is supported by a 500 $\mu$m-thick and 5 mm-square silicon frame. On its non-suspended part, 200 $\mu$m away from the membrane, two 750 $\mu$m-wide and 500 nm-high rectangular dielectric spacers were deposited. The 55 nm-thick membrane is supported by a 500 $\mu$m-thick and $4\times 5$ mm silicon frame. Both membranes were assembled parallel to each other into a monolithic sandwich while monitoring its transmission under broadband illumination and gluing the sides of the spacers using a zero-expansion, UV curing resist (OrmoCom Clear FX)~\cite{Nair2017,Naserbakht2019}. The optical transmission of the assembled sandwich, shown in Fig.~\ref{fig:trans}(c), is consistent with that of a planar Fabry-Perot cavity with a gap of 501(1) nm. The high quality of the interference fringe contrast also indicates good parallelism between the two membranes.

\begin{figure}
\begin{subfigure}{0.49\columnwidth}
\centering
\includegraphics[width=0.9\columnwidth]{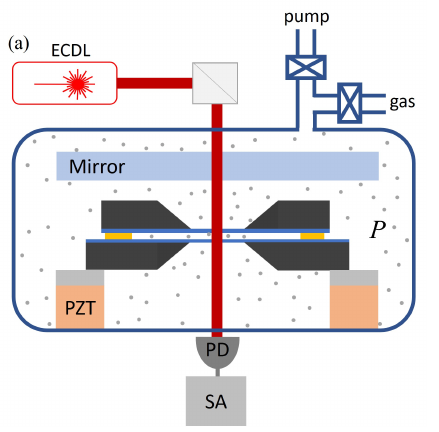}\\
\includegraphics[width=\columnwidth]{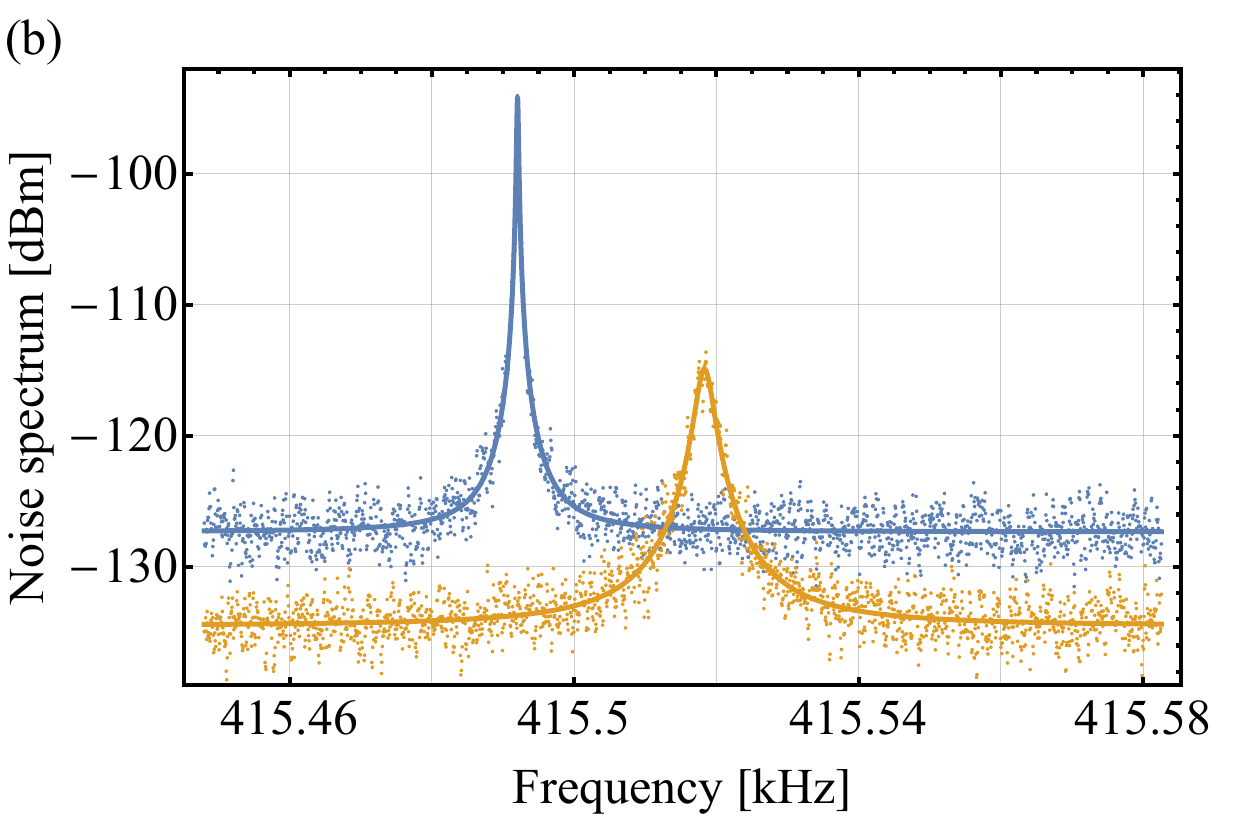}
\end{subfigure}
\begin{subfigure}{0.49\columnwidth}
\centering
\includegraphics[width=1\columnwidth]{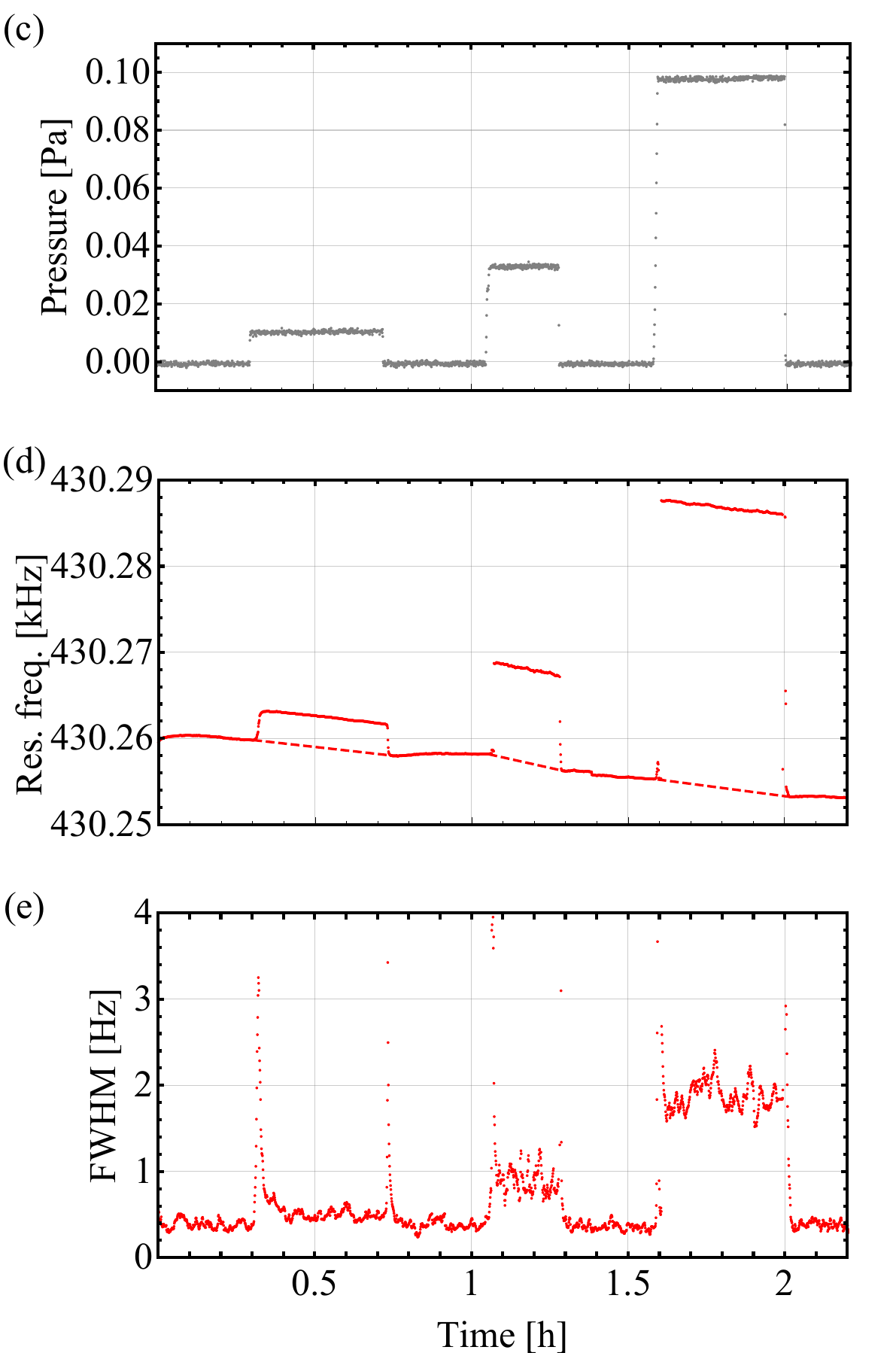}
\end{subfigure}
\caption{(a) Schematic of the setup used for the pressure measurements. (b) Thermal noise spectra of the low frequency mode in vacuum (blue) and for a nitrogen pressure of 0.3 Pa (yellow). A 10 dBm vertical offset has been added for clarity to the vacuum noise spectrum. The lines show the results of fits to the data with a Lorentzian function, whose center yield the mechanical resonance frequency and whose FWHM yields the damping rate $\gamma/(2\pi)$.  Example of a measurement sequence with 3 pressure steps. (c) Evolution in time of the pressure measured by the commercial capacitive drum sensor. (d) Corresponding evolution of the high frequency mode resonance frequency. The dashed line indicate the extrapolated vacuum resonance frequency during the pressure steps. (e) Corresponding evolution of the FWHM of the thermal noise spectrum.}
\label{fig:setup}
\end{figure}

The membrane sandwich is placed in the main vacuum chamber of a multi-chamber system otherwise designed for the measurement of saturation vapor pressures of low-volatile substances~\cite{Nielsen2023}. The chamber is actively temperature stabilized and thermally isolated in order to minimize in particular mechanical frequency drifts due to external temperature fluctuations. The vibrations of the membranes are measured interferometrically by monitoring the transmission of monochromatic light (900 nm) from an external cavity diode laser through a Fabry-Perot interferometer consisting of a 50\% transmitting mirror and the membrane sandwich (Fig.~\ref{fig:setup}(a)). The light intensity injected into the interferometer is actively stabilized using an acoustooptic modulator to minimize laser power fluctuations-related mechanical frequency drifts. The light transmitted by the interferometer is collected by a low-noise transimpedance photodectector and the resulting photocurrent is fed to a low resolution bandwidth spectrum analyzer. This provides us with the thermal noise spectrum of the membrane modes. In this work we focus on the membranes' fundamental (lowest frequency) drummodes, which give the strongest squeeze film pressure response.  In vacuum these modes have resonance frequencies of 430 kHz and 415 kHz, respectively, with very similar Qs of $1.2\times 10^6$ and $1.3\times 10^6$.

In Fig.~\ref{fig:setup}(b) are representative examples of recorded spectra showing the thermal noise spectra of the fundamental mode of low frequency mode (LF) membrane in high vacuum and at a nitrogen pressure of 0.3 Pa. The gas-induced positive resonance frequency shift of $\sim 26$ Hz and broadening of the spectrum (corresponding here to an increase in the damping rate by a factor $\sim 10$) are clearly visible.


The pressure responsivity of the sensors was evaluated by monitoring the pressure in the chamber during a succession of steps involving rapidly (within tens of seconds) leaking in nitrogen to achieve a given static pressure in the chamber (in absence of pumping) and then pumping the chamber down again to high vacuum ($<10^{-5}$ Pa) to keep track of the drift of the vacuum resonance frequency of the mode considered due to small temperature variations. The pressure in the chamber is monitored by two absolute capacitive diaphragm sensors---BCEL7045 0.1 mbar (Edwards) for the range $1\times 10^{-3}$ to $10$ Pa and CDG045D 10 mbar (Inficon) for the range $5\times 10^{-1}$ to 1000 Pa---with specified accuracies of 0.15\% and resolutions of 0.003\%. The resonance frequency shift at a given pressure can then be obtained by subtraction of the measured resonance frequency at a given time during the pressure step by the extrapolated vacuum resonance frequency at the same time, as shown in Fig.~\ref{fig:setup}(d). Likewise, the damping of the mode can be evaluated by measuring the FWHM of the thermal noise spectrum in frequency space, which is related to the damping rate $\gamma$ by FHWM=$\gamma/(2\pi)$. In addition to be more noisy due to the lower responsivity in terms of pressure broadening, the damping is also observed to strongly vary both at the beginning and at the end of the pressure steps (Fig.~\ref{fig:setup}(e)). We surmize that a temporary modification of the stress and/or clamping of the membranes  follows the sudden application of the pressure. To avoid introducing potential systematic errors due to these effects only points away from the edges of the pressure steps are used for the analysis of the frequency shift and damping at a given pressure.

\begin{figure}
\begin{subfigure}{0.49\columnwidth}
\centering
\includegraphics[width=\columnwidth]{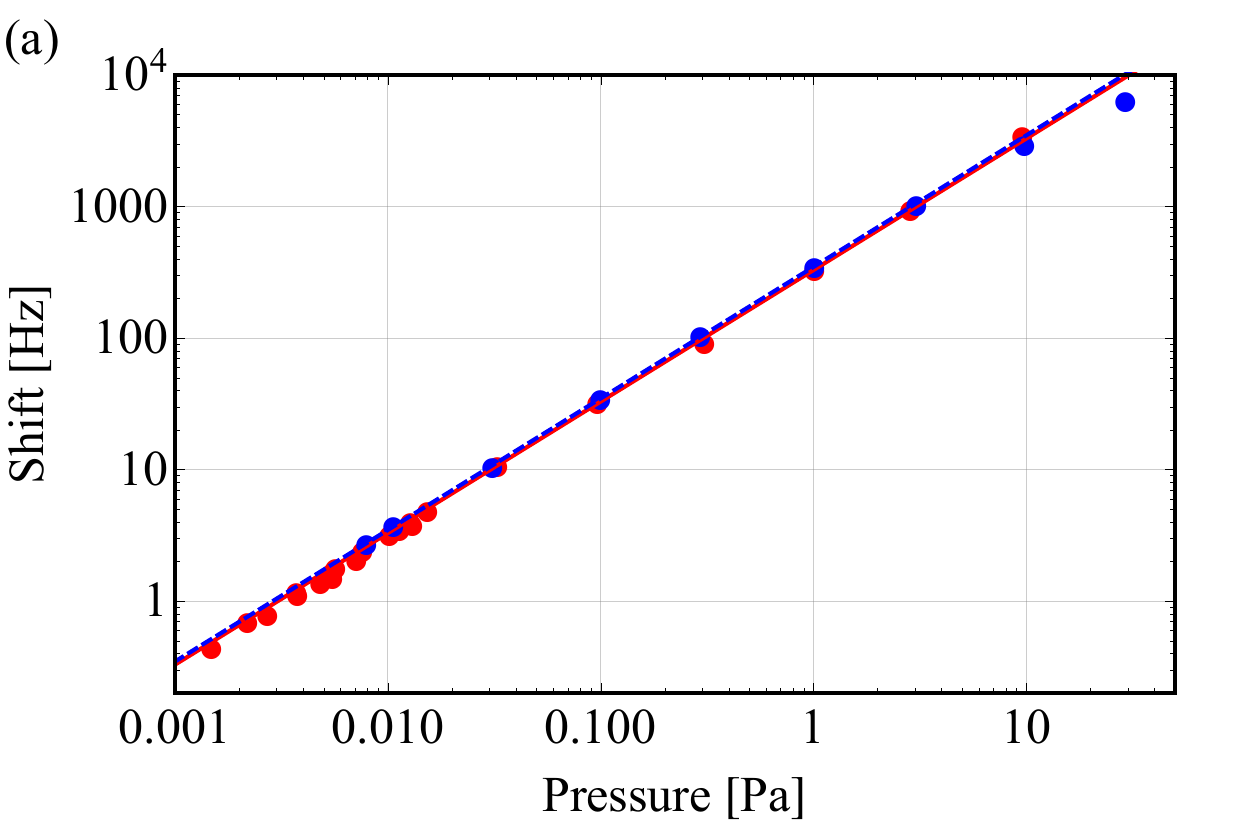}
\end{subfigure}
\begin{subfigure}{0.49\columnwidth}
\centering
\includegraphics[width=\columnwidth]{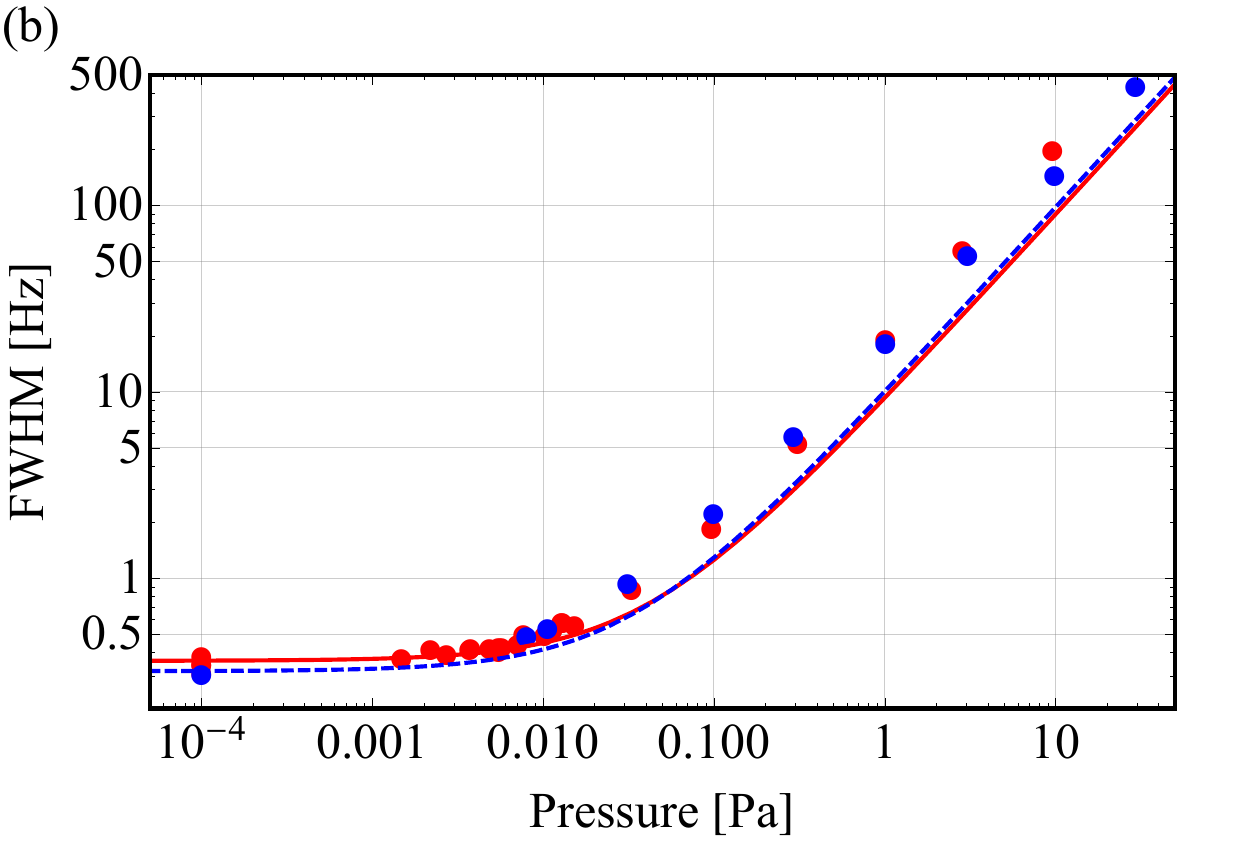}
\end{subfigure}
\caption{(a) Mechanical resonance frequency shift for the high (red) and low (blue) frequency modes as a function of pressure. The full and dashed lines show the results of linear fits with slopes of 328(8) Hz/Pa and 347(9) Hz/Pa, respectively. (b) Corresponding FWHM of the thermal noise spectrum, where the FWHM in frequency is related to the damping rate by FWHM=$\gamma/2\pi$. The full and dashed lines show the sum of the intrinsic (vacuum) damping and of the kinetic damping predictions of Eq.~(\ref{eq:gammakin}).}
\label{fig:responsivity}
\end{figure}

The variations of the resonance frequency shift and the FWHM of the thermal noise spectra of the fundamental modes of both membranes, averaged over each step (the vicinity of the edges notwithstanding), are shown in the range $10^{-4}$ to 20 Pa in Figs.~\ref{fig:responsivity}(a) and (b), respectively. Linear fits to the measured resonance frequency shifts yield responsivities of 328(8) and 340(1) Hz/Pa, respectively, which represents a 10-fold improvement with respect to the sandwiches of Ref.~\cite{Naserbakht2019}. These values are also consistent with those expected from Eq.~(\ref{eq:shift}), assuming a density of 3200 kg/m$^3$. Given the relatively large difference ($\sim 15$ kHz) in the vacuum resonance frequencies as compared to the gas-induced frequency shifts, the oscillations of both modes remain essentially uncorrelated in the pressure range considered, although the slightly smaller shift observed at the highest pressure for the low frequency mode indicates the onset of mode hybridization, as was observed in Refs.~\cite{Naesby2017,Naserbakht2019}.

The solid lines in Fig.~\ref{fig:responsivity}(b) show the sum of the intrinsic damping in vacuum, due to clamping/material losses, and of the kinetic damping predictions of Eq.~(\ref{eq:gammakin}), which yields a broadening of 9-10 Hz/Pa for the assumed value of the density, substantially lower than the observed 18-19 Hz/Pa. This suggests the presence of significant squeeze film damping, as will be discussed in more detail later.

\begin{figure}
\center\includegraphics[width=0.6\columnwidth]{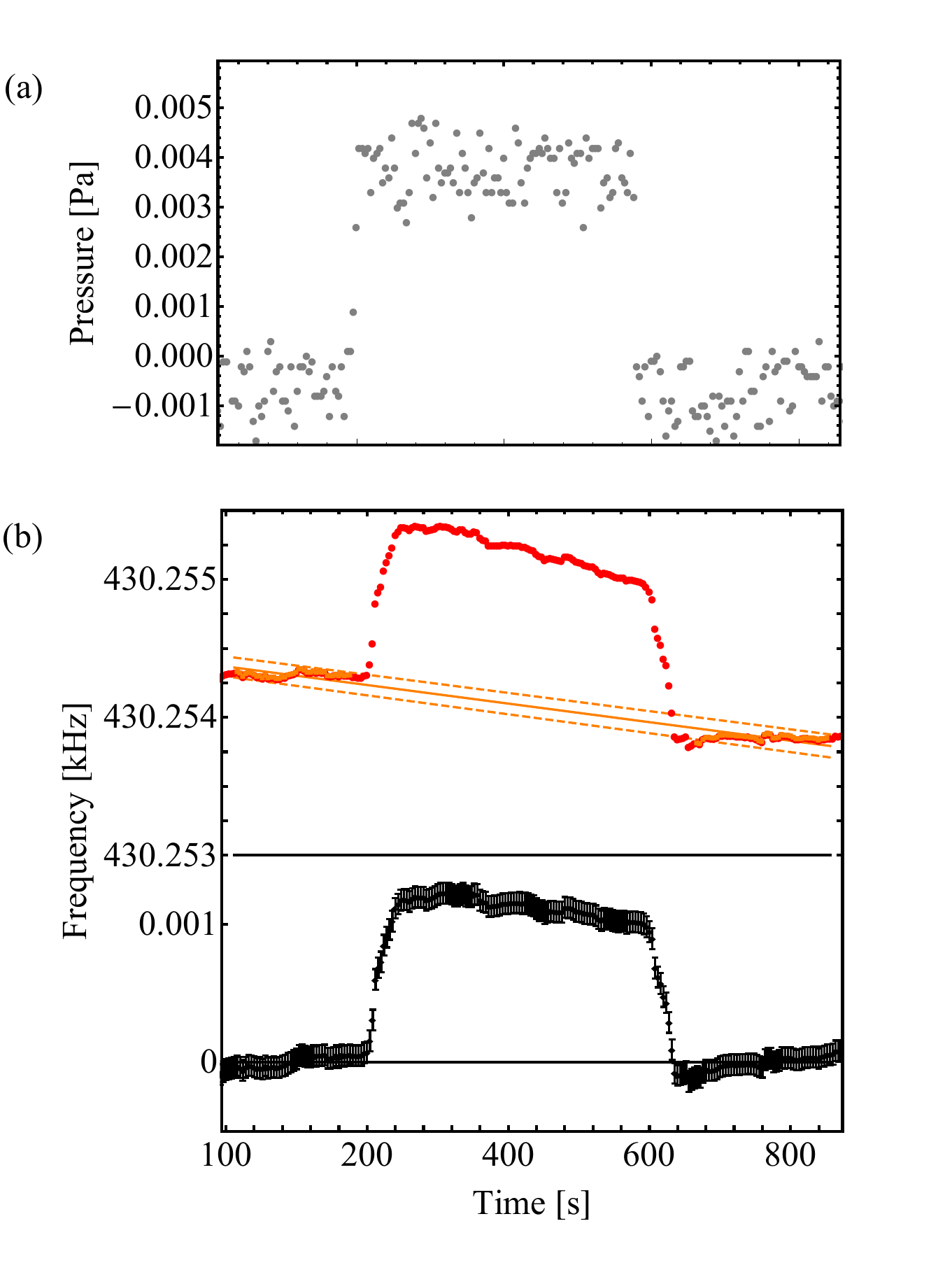}
\caption{(a) Evolution in time of the pressure measured by the commercial absolute capacitive sensor during a 3.5 mPa pressure step. (b) Corresponding evolution of the high frequency mode resonance frequency. The orange line shows the results of a linear baseline interpolation of the mechanical resonance frequency in vacuum, the dashed lines indicating the $\pm 1\sigma$ standard deviations. The uncertainties in the individual mechanical frequency measurements associated with the results of the Lorentzian fits to the spectra are typically smaller than the point size. The shifts inferred after baseline substraction are shown as the black points, offset for clarity.}
\label{fig:sensitivity}
\end{figure}

Figure~\ref{fig:sensitivity} shows the evolution of the mechanical resonance frequency of the high frequency mode during a pressure step from high vacuum to $\sim 3.5\times 10^{-3}$ Pa, as measured by the commercial absolute capacitive sensor. To evaluate the uncertainty in the resonance frequency shift during such a step a linear interpolation of the resonance frequency in vacuum before and after the step is performed, whose result together with its $\pm 1\sigma$ standard deviations are shown as the orange lines in Fig.~\ref{fig:sensitivity}(b). The average shift measured during this step is found to be $1.070\pm 0.075$ Hz, dominated by the uncertainty associated with baseline correction. While the sensitivity and accurary of the current commercial sensor does not allow for a direct evaluation of the sensitivity of the squeeze film frequency shift measurement, one can estimate it in the typical current experimental conditions by dividing the 0.075 Hz shift uncertainty by the responsivity of 328 Hz/Pa, which yields a 0.2 mPa sensitivity. Further improvements in sensitivity can be envisaged by improving the temperature control and accuracy of the baseline corrections, as well as by using higher Q, lower frequency or even thinner membranes~\cite{Reinhardt2023}.

\begin{figure}
\includegraphics[width=0.5\columnwidth]{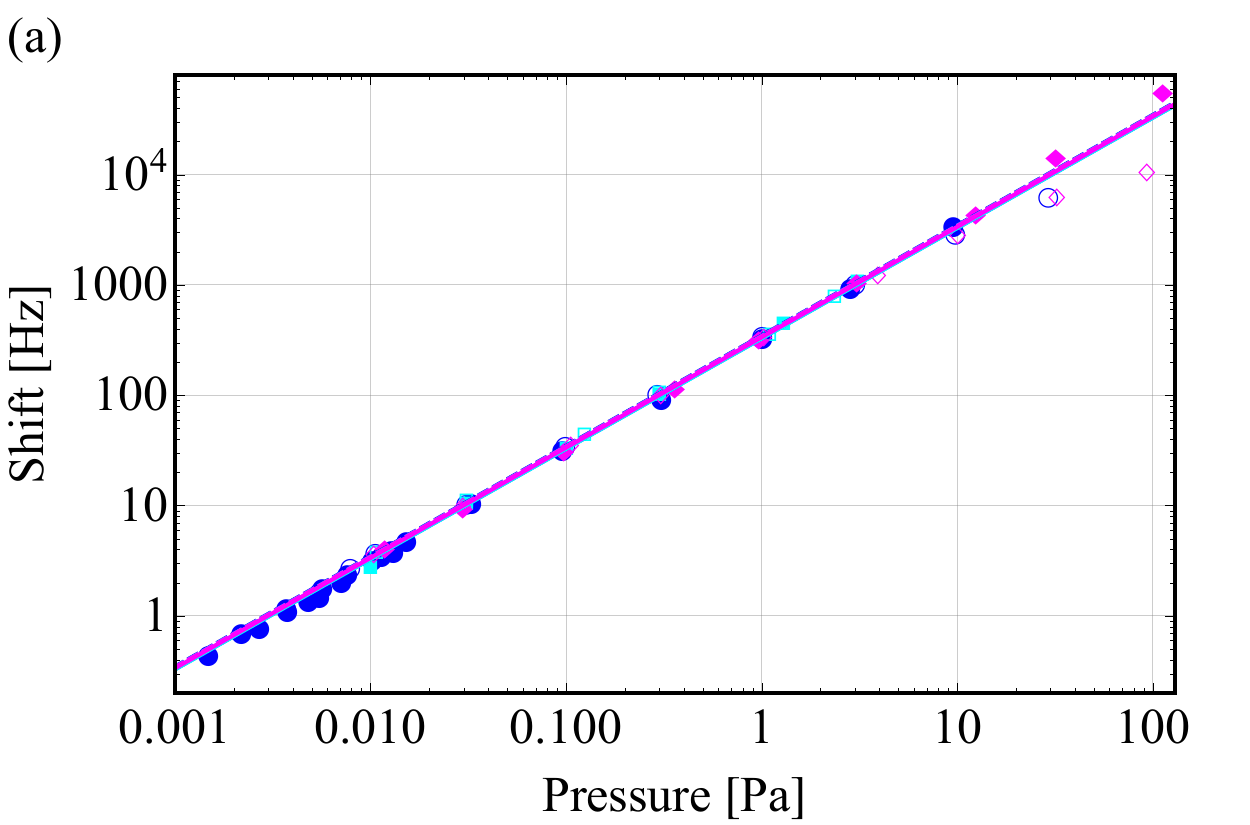}\\
\includegraphics[width=0.5\columnwidth]{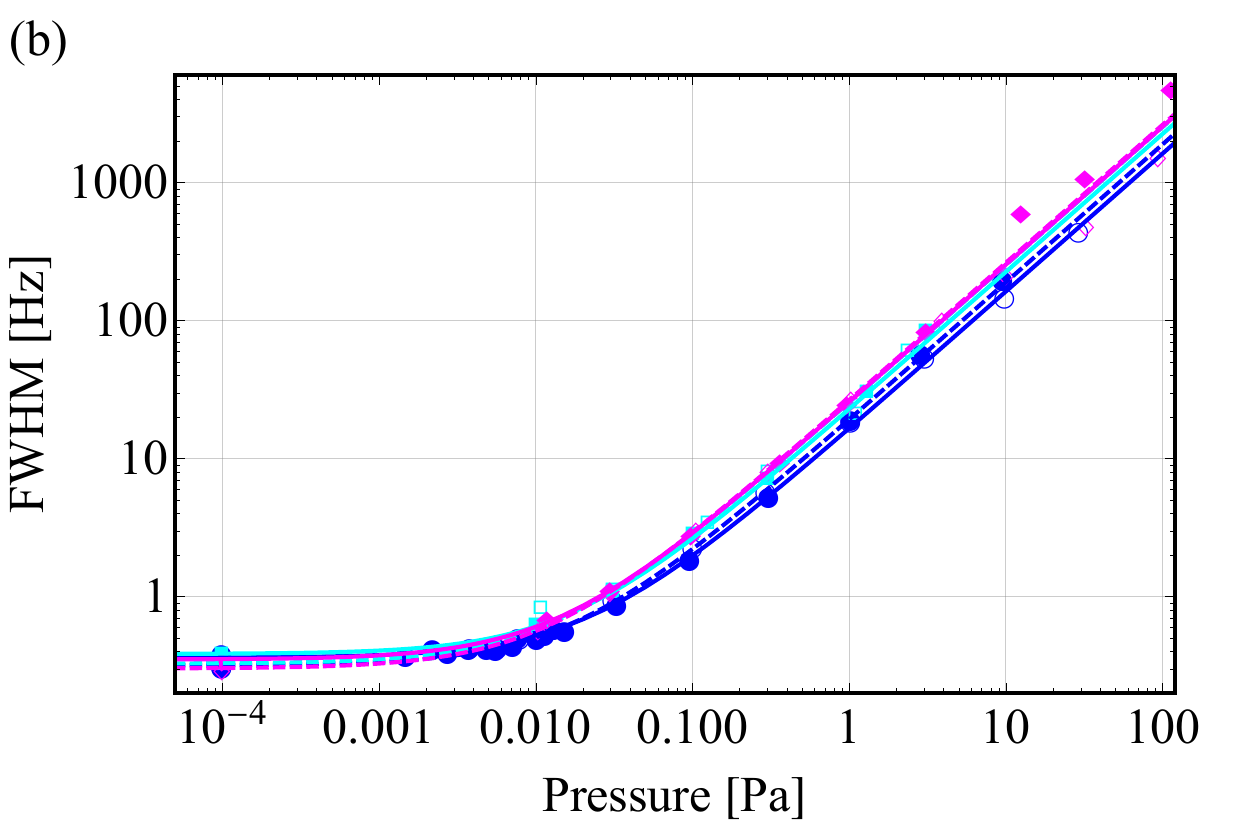}\\
\includegraphics[width=0.5\columnwidth]{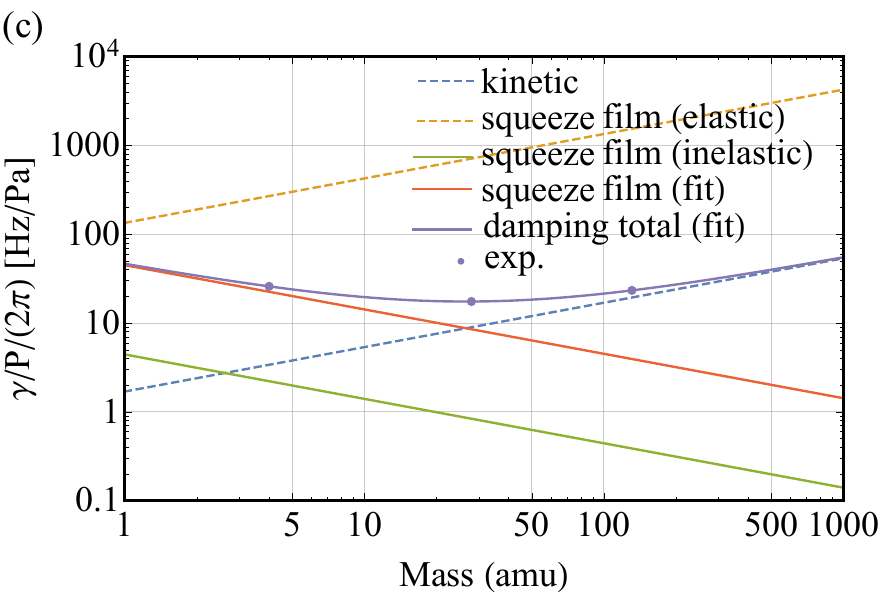}
\caption{Mechanical resonance frequency shift (a) and thermal noise spectrum FWHM (b) as a function of pressure for different gases: N$_2$ (circles), He (squares) and Xe (diamonds), the full and empty symbols showing the results for the high and low frequency modes, respectively. The plain and dashed lines show the result of linear fits for both modes (see text). (c) Variation with the mass of the molecules of the various damping contributions (in Hz/Pa) for the high frequency mode.}
\label{fig:gasses}
\end{figure}

To assess the absolute character of the pressure response, similar experiments were performed leaking helium ($M=4$) and xenon ($M=131$) into the chamber. The observed squeeze film-induced frequency shifts, shown in Fig.~\ref{fig:gasses}, are essentially identical for all gases, thus supporting the species-independent pressure response of the sensor. For completeness, the FWHM of the thermal noise spectra are also shown in Fig.~\ref{fig:gasses}(b), together with linear fits of the form $\gamma_0+\gamma_P P$, with the measured vacuum damping $\gamma_0$ being a fixed parameter (see also Supplementary material). The inferred responsitivities, both in terms of frequency shifts and noise spectrum broadening for both modes, are reported for all three gases in Table~\ref{tab}. The frequency shift responsivities are at least one order of magnitude larger than the damping responsivities.

\begin{table}
  \caption{Measured responsivities in terms of resonance frequency shift $\Delta f/P$ and pressure broadening $\gamma_P$ (in Hz/Pa) of both high (HF) and low (LF) frequency modes for N$_2$, He and Xe. The 3$^\textrm{rd}$ column gives the kinetic damping broadenings (in Hz/Pa) expected from Eq.~(\ref{eq:gammakin}). The 4$^\textrm{th}$ and 5$^\textrm{th}$ columns give the squeeze film damping contribution expected from fully elastic and fully inelastic collision models, respectively. The 6$^\textrm{th}$ column shows the squeeze film damping contribution resulting from a fit to the data with an equalization time $\tau=\epsilon\tau^\textrm{Sui}$ in a partially elastic collision model. The last column shows the broadening due to the total damping $\gamma_\textrm{tot}=\gamma_\textrm{kin}+\gamma_\textrm{sq}^\textrm{fit}$.}
  \label{tab}
  \begin{tabular}{|c||cc||cc||cc|cc|cc|cc||cc|}
    \hline
    \multirow{2}{*}{Gas} &
      \multicolumn{2}{c||}{$\Delta f/P$} &
      \multicolumn{2}{c||}{$\gamma_P$} &
      \multicolumn{2}{c|}{$\gamma_\textrm{kin}/P/(2\pi)$} &
      \multicolumn{2}{c|}{$\gamma_\textrm{sq}^\textrm{Bao}/P/(2\pi)$} &
      \multicolumn{2}{c|}{$\gamma_\textrm{sq}^\textrm{Sui}/P/(2\pi)$} &
      \multicolumn{2}{c||}{$\gamma_\textrm{sq}^\textrm{fit}/P/(2\pi)$} &
      \multicolumn{2}{c|}{$\gamma_\textrm{tot}/P/(2\pi)$} \\
   & HF & LF & HF & LF & HF & LF & HF & LF & HF & LF & HF & LF & HF & LF\\
    \hline
    He      &  328 & 343 & 22.5 & 25.1 & 3.4 & 3.7 & 268 & 292 & 2.2 & 2.6 & 22.4 & 26.2 & 25.8 & 29.9\\
    N$_2$ & 329 & 347 & 17.5 & 19.0 & 8.9  & 9.7  & 709 & 773 & 0.8 & 1.0 & 8.5 & 9.9 & 17.4 & 19.7\\
    Xe      &  337 & 342 & 25.4 & 26.1 & 19.3 & 21.1 & 1533 & 1672 & 0.4 & 0.5 & 3.9 & 4.6 & 23.2 & 25.6\\
    \hline
  \end{tabular}
\end{table}

While the magnitude of the damping observed for xenon is close to that of the expected kinetic damping, the dampings observed for both helium and nitrogen are substantially higher than the predicted kinetic damping, which suggests the presence of significant squeeze film damping. Given the canonical situation considered (large area membrane, small gap and rarefied gas regime) one can relatively straightforwardly attempt to account for the squeeze film damping contribution on the basis of several models, which distinguish themselves by the nature of the collisions between the gas molecules and the membranes. In their resonant energy transfer model, which assumes perfectly elastic collisions, Bao {\it et al.}~\cite{Bao2007} predict a squeeze film damping of 
\begin{equation}
\gamma_\textrm{sq}^\textrm{Bao}=\frac{4a}{\pi^2\bar{v}d}\frac{P}{\rho t},
\label{eqSM:gammasqBao}
\end{equation}
 $\bar{v}=\sqrt{8RT/\pi M}$ is the mean thermal velocity of the gas molecules at temperature $T$. The magnitude of $\gamma_\textrm{sq}^\textrm{Bao}$ is reported in Table~\ref{tab} for both modes and the three gases. In contrast, Suijlen and coworkers assume totally inelastic collisions~\cite{Suijlen2009} and derive an analytical expression for the equalization time $\tau$, given in the geometry and regime of interest by
\begin{equation}
\tau^\textrm{Sui}=\frac{8a^2}{\pi^3 d \bar{v}}.
\label{eq:tauS}
\end{equation} This analytical expression was verified through simulations and experiments in the inelastic squeeze film force regime~\cite{Suijlen2009}. It shows in particular that heavier molecules/atoms are trapped for a longer time in the gap and their squeeze film damping contributions is comparatively smaller than for lighter molecules. This is in contrast with the kinetic damping of Eq.~(\ref{eq:gammakin}),
which increases with $\sqrt{M}$. Using Eq.~(\ref{eq:gammasq}) yields predictions for the squeeze film damping rate in the inelastic collision regime equal to $\gamma_\textrm{sq}^\textrm{Sui}$, which are also reported in Table~\ref{tab}. Looking at Table~\ref{tab} it is clear that assuming perfectly elastic collisions clearly overestimates the magnitude of the squeeze film damping, while the inelastic collisions model underestimates it. Partially elastic collisions, resulting in shorter trapping times, would thus represent a reasonable explanation. In a simple attempt to account for the experimentally observed damping rates for helium, nitrogen and xenon, we assume an equalization time of the form~(\ref{eq:tauS}), $\tau=\epsilon \tau^\textrm{Sui}$, 
but scaled by a factor $\epsilon$ to take the faster escape of the molecules due to partially elastic collisions into account, and perform a global fit to the observed dampings for all modes and gases. We restrict the fitting range to pressures below 1 Pa to exclude potential mode hybridization effects. This yields a value of $\epsilon=0.098(4)$, which suggests a equalization time shorter by about one order of magnitude than that predicted by the full accomodation model of Ref.~\cite{Suijlen2009}. The variation of the different pressure broadenings with the mass of the molecules are shown in Fig.~\ref{fig:gasses}(c). Note that the resulting Weissenberg numbers are $W_i\sim270$, 720, 1550 for He, N$_2$ and Xe, respectively, consistently with an essentially elastic squeeze film force. On the one hand, these observations call for further investigations of squeeze film dynamics in such small gap structures. On the other hand, they highlight the fact that species-dependent effects are in general complex and have to be carefully assessed in damping-based determinations of pressure.

To conclude we reported on the assembly and characterization of a SiN membrane sandwich-based squeeze film pressure sensor, which displays record responsivity in terms of squeeze film induced-resonance frequency shift of >300 Hz/Pa and sensitivity below the millipascal level, thus surpassing the best commercial capacitive drum sensors. The demonstrated species-independent squeeze film frequency shift provides a robust and direct dtermination of the absolute pressure. As mentioned above, achieving better sensitivity can be envisaged by improving the temperature control and accuracy of the baseline corrections, as well as by using higher Q, lower frequency or even thinner membranes. Another interesting avenue to explore in the future is the hybridized regime, in which the vibrational modes of both membranes become strongly coupled by the gas, potentially leading to stronger squeeze film shift responsivity and a modified damping response, the onset of which may be surmized by looking at the highest pressure datapoints in Fig.~\ref{fig:gasses}. While the modes of the sandwich used in this work were too far from frequency degeneracy to thoroughly investigate this regime, it should be possible to explore these squeeze film dynamics using more mechanically identical membranes~\cite{Naesby2017,Naserbakht2019} or by actively tuning their mode spectrum, e.g. using temperature~\cite{StGelais2019} or stress control~\cite{Naserbakht2019a,Naserbakht2019b}. Yet another prospect for increased pressure sensitivity would be the use of highly reflective membranes~\cite{Parthenopoulos2021,Darki2022} sandwiches to form high-finesse flexible microcavities~\cite{AlSumaidae2021}, which may provide more efficient displacement readout sensitivity.

\begin{acknowledgement}

The authors thank the Independent Research Fund Denmark and the Villum Foundation for financial support. We are also grateful to Norcada for their assistance with the design and fabrication of the samples.

\end{acknowledgement}

\begin{suppinfo}


\begin{figure}
\includegraphics[width=\columnwidth]{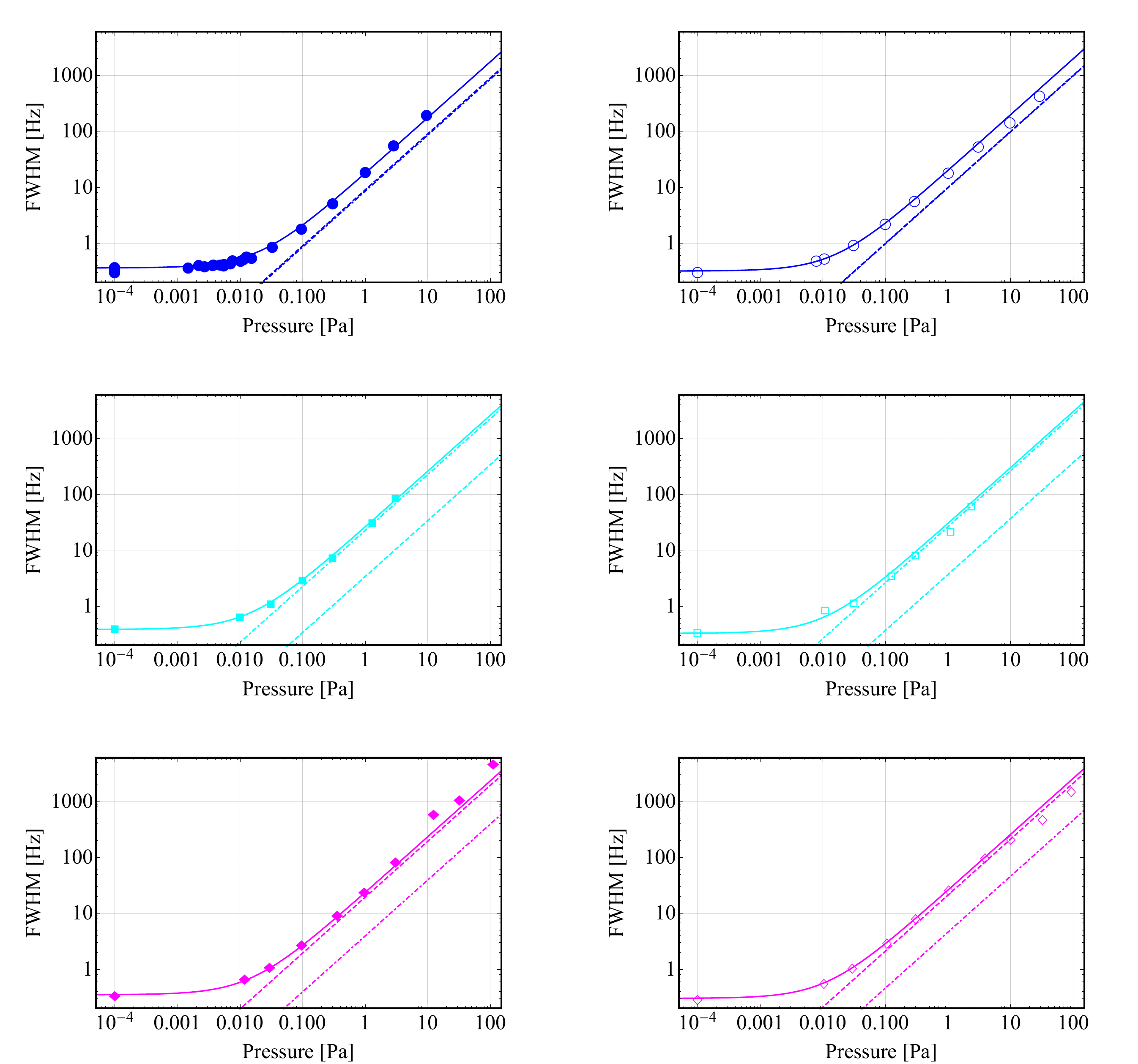}\\
\caption{FWHM as a function of pressure for the high (left column) and low (right column) frequency modes for N$_2$ (top row), He (middle row) and Xe (bottom row). The dashed lines show the kinetic damping contributions, the dot-dashed line the squeeze film damping contribution (partially elastic collision model) and the plain line the total damping.}
\label{fig:gasses}
\end{figure}

\end{suppinfo}

\bibliography{sandwich_bibtex}

\end{document}






\section{Damping analysis}

\begin{figure}
\includegraphics[width=\columnwidth]{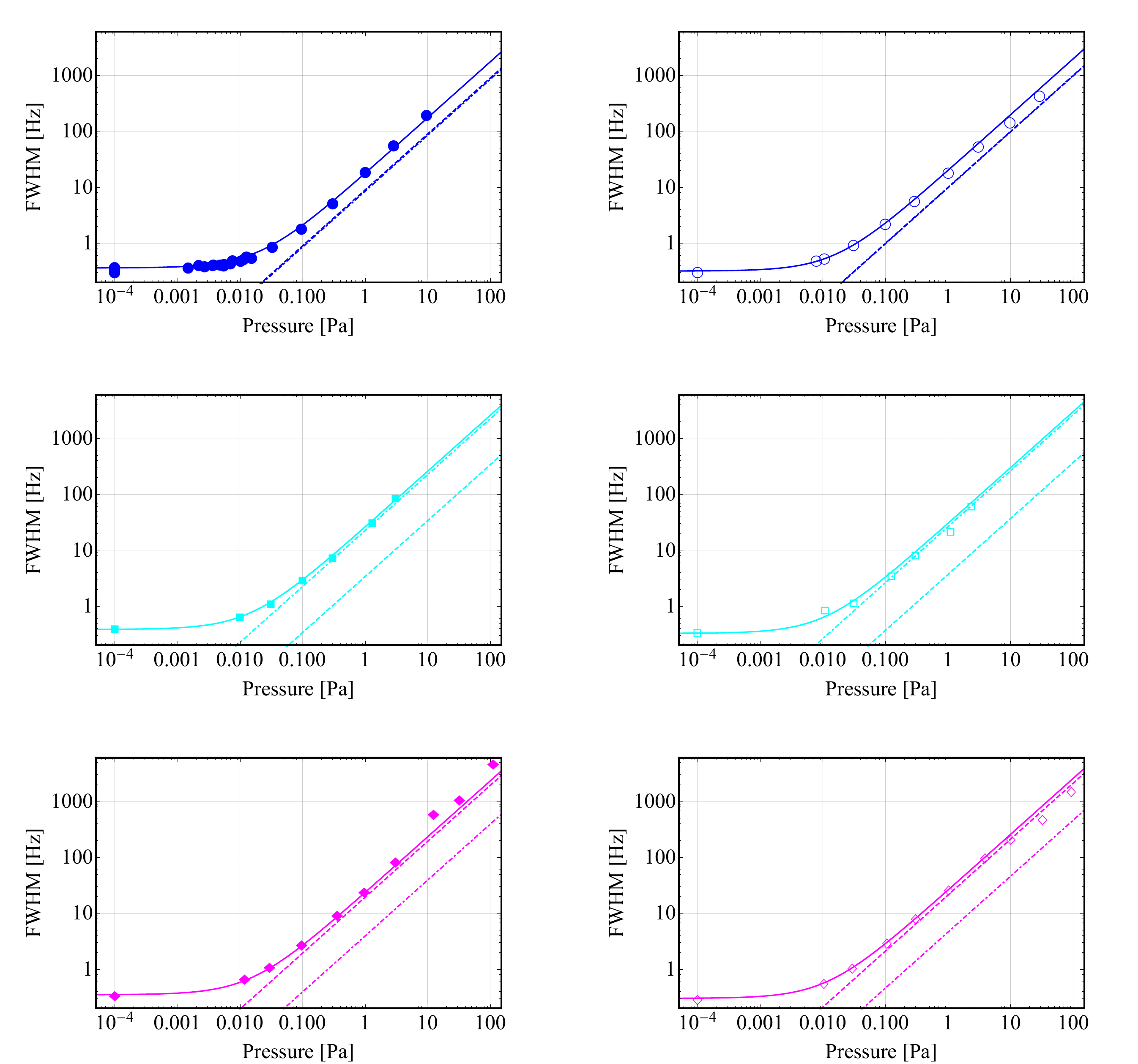}\\
\caption{FWHM as a function of pressure for the high (left column) and low (right column) frequency modes for N$_2$ (top row), He (middle row) and Xe (bottom row). The dashed lines show the kinetic damping contributions, the dot-dashed line the squeeze film damping contribution (partially elastic collision model) and the plain line the total damping.}
\label{fig:gassesSI}
\end{figure}

Figure~\ref{fig:gassesSI} shows the variations with pressure of the measured FWHM of the thermal noise spectrum for both modes and for the three gasses (Fig.~5(b) of the main manuscript), as well as the respective contributions of kinetic and squeeze film damping in a partially elastic collision model, as detailed in the main manuscript.

%
%
%
%
%

\bibliography{sandwich_bibtex}